\documentclass[review]{elsarticle}

\usepackage{lineno,hyperref}
\usepackage{graphicx}
\usepackage[american]{babel}
\usepackage{amsmath}
\usepackage{amsthm}
\usepackage{amsfonts}
\usepackage{amssymb}
\usepackage{color}
\newtheorem{Def}{Definition}
\newtheorem{Proposition}[Def]{Proposition}

\usepackage{ragged2e}
\usepackage{lipsum}
\usepackage{enumerate}
\modulolinenumbers[5]

\journal{Statistical papers}

\begin{document}

\begin{frontmatter}

\title{Indexed Markov Chains for financial data: testing for the number of states of the index process} 

\author[gda]{Guglielmo D'Amico}
\address[gda]{Dipartimento di Farmacia, Universit\`a "G. D'Annunzio" di Chieti-Pescara,  66013 Chieti, Italy}
\author[al]{Ada Lika} 
\address[al]{Dipartimento di Scienze Economiche ed Aziendali, 
	Universit\`a degli studi di Cagliari, 09123 Cagliari, Italy}
\author[al]{Filippo Petroni}

\begin{abstract}
A new branch based on Markov processes is developing in the recent literature of financial time series modeling. In this paper, an Indexed Markov Chain has been used to model high frequency price returns of quoted firms. The peculiarity of this type of model is that through the introduction of an Index process it is possible to consider the market volatility endogenously and two very important stylized facts of financial time series can be taken into account: long memory and volatility clustering.
In this paper, first we propose a method for the optimal determination of the state space of the Index process which is based on a change-point approach for Markov chains. Furthermore we provide an explicit formula for the probability distribution function of the first change of state of the index process.
Results are illustrated with an application to intra-day prices of a quoted Italian firm from January $1^{st}$, 2007 to December $31^{st}$ 2010.
\end{abstract}


\end{frontmatter}

\section{Introduction}
\label{intro}

Markov chains and semi-Markov models have been used in a vast variety of disciplines such as physics and engineering, see, e.g. \cite{brem99,limnios2001semi}; finance and actuarial sciences are not an exception, see e.g. \cite{kijima2013stochastic,janssen2006applied}. Their success in many applicative domains probably resides from one side in the simplicity of the model definition and basic generating idea and from the other side in the unnecessary specification of particular parametrization of the model.\\
\indent However, Markov chains models have not been used in modeling intra-day dynamics of financial returns, a notable exception is given by \cite{shiy99}. One of the main inadequacies of a Markov chain model in the description of intra-day data is the low autocorrelation function of the square of returns which in turn means poor reproduction of the volatility clustering, see \cite{d2012semi}. This problem has stimulated the research of plausible solutions that were not in the practical unmanageable direction of increasing the order of the Markov chain. In a series of papers, the idea of indexed semi-Markov processes was advanced, see \cite{dami11amm,d2011semi,d2012weighted,d2013multivariate,d2017Copula}. These processes revealed to be useful to reproduce important stylized fact of high-frequency financial data as the absence of autocorrelations in returns, the gain/loss asymmetry, autocorrelation function of the squares of returns as well as multivariate extensions useful to model financial portfolios.
These indexed models have a very appealing property: they are endogenous models in the sense that there is no need to introduce noise process transformations like in the ARCH/GARCH methodology neither to introduce non observables processes (often Hidden Markov chains) to be able to incorporate volatility regimes, see e.g. \cite{bulla06,augu14,holz16}.\\
\indent The index process is a moving average of a function of past returns and in general it assumes values in the real set of numbers. For the practical application of the indexed model it is a crucial point the discretization of the state space of the index process. In the previous articles this step was executed in an heuristic way considering the histogram of the index process and considering a partition of the distribution in a given number of sub-intervals. Each state in the application correspond to a level of volatility that changes the dynamic of the returns. In this paper we focus on Indexed Markov Chain (IMC) models and we proceed to develop a statistical procedure that permits the identification of the optimal partition of the state space of the index process and the consequential identification of the necessary volatility regimes to be considered for an accurate modeling. In this way we obtain an automatized procedure for the implementation of the model. The procedure is an adaptation of the change-point techniques for Markov chains as developed by \cite{polansky2007detecting}. In \cite{polansky2007detecting} statistical techniques were used to determine the times where a change of dynamic of the Markov chain occur and a different transition probability matrix should be applied. In our paper we are interested in determining the threshold values of the index process where the process of returns undergoes a significant change depending on the index process (volatility), marking a discontinuity with the past. We identify the number and size of the thresholds and consequently we estimate one transition probability matrix for each value of the index. Furthermore, the problem of determining the probability distribution function of the next change in the index process (volatility) is addressed and an explicit formula is determined. The developed concept are applied to real intra-day financial data concerning the stock ENI traded at {\em Borsa Italiana} (the Italian Stock Exchange) from  January $1^{st}$, 2007 to December $31^{st}$ 2010. The empirical results shows that the optimal number of change-point is four and consequently we identify five different levels of volatility that are necessary to obtain results close to real data.\\
\indent  The paper is organized as follows: in Section \ref{sec:imc} a general description of Markov models where an Index process is introduced is provided, in Section \ref{sec:chp} the discretization of the state space of the Index process based on the change-point approach is formalized and an explicit formula for the calculation of the probability of the process to step in one of the Index levels is shown in Section \ref{sec:predict}. Lastly, in Section \ref{sec:emp} empirical applications of the methodology are shown.  

\section{Indexed Markov processes}
\label{sec:imc}

Since the application of the change point approach for the discretization of the state space is based on the peculiarities of the Indexed Markov models, it seems useful to briefly summarize the main characteristics of this model.\\
\indent On a complete probability space $(\Omega, \mathcal{F},\mathbb{P})$ we define a sequence of random variables $\{S_{n}\}_{n\in \mathbb{N}}$  denoting the price of a financial asset at time $n\in \mathbb{N}$. We concentrate our attention on the log-return process, i.e. on the process $R_{n}$ defined by $R_{n}=\log(S_{n}/S_{n-1})$.  According to \cite{d2013multivariate}, the log-returns are transformed into a sequence of discrete returns by means of a map: 
\begin{equation}
\label{map}
\mathcal{M}:\mathbb{R} \longrightarrow E=\{-z_{min}\Delta, ..., -\Delta, 0, \Delta, ..., z_{max}\Delta\},
\end{equation}
where $\Delta$ is the grid amplitude of $E$.\\
\indent We assume that the discrete return $J_{n}:=\mathcal{M}(R_{n})$ attains the value $i\Delta$ whenever the continuous return $R_{n}$ belongs to the interval $\bigg((i-\frac {1}{2})\Delta , (i+\frac{1}{2})\Delta \bigg]$. The lowest discrete return $J_{n}=-z_{min}\Delta$ is achieved when the continuous return $R_{n}\leq (-z_{min}+\frac {1}{2})\Delta$ whereas the highest discrete return $J_{n}=z_{max}\Delta$ is assigned whenever $R_{n}> (z_{max}-\frac {1}{2})\Delta$.\\
\indent We also consider the stochastic process $\{V_{n}^{m}\}_{n\in \mathbb{N}}$ with values in $\mathbb{R}$. The random variable $V_{n}^{m}$ describes the value of the index process at the $n$-th transition and is defined as follows:
\begin{equation}
\label{index}
V_{n}^{m}=\dfrac{\sum_{k=0}^{m-1} f(J_{n-k})}{m}.	
\end{equation}
\indent Formula $(\ref{index})$ expresses the index process as a moving average of a function $f$ of past returns $(J_{n},J_{n-1},\ldots,J_{n-m+1})$.  As time progresses from time $n$ to time $n+1$, the new return $J_{n+1}$ substitutes the return $J_{n-m+1}$ and a new value of the index $V_{n+1}^{m}$ is determined. The parameter $m$ is called the memory of the process and should be calibrated to data. In the applicative section we will describe the calibration of $m$ and the choice of the function $f$ in a way that they can contribute to the detection of the volatility.\\
\indent The Indexed Markov chain model is defined imposing a probabilistic relationship between the random processes $J_{n}$ and $V_{n}^{m}$:
\begin{equation}
\begin{aligned}
& \mathbb{P}(J_{n+1}=j|J_{n}=i,V_{n}^{m}=v, \sigma(J_{h},V_{h}^{m},\,h=0,\ldots,n))\\
&= \mathbb{P}(J_{n+1}=j|J_{n}=i,V_{n}^{m}=v)=: p_{ij}(v), 
\end{aligned}
\label{hp}	
\end{equation}
where $\sigma(J_{h},V_{h}^{m},\,h=0,\ldots,n)$ is the natural filtration of the bi-variate process. 

The Eq. \eqref{hp} states that the value of the price return process at the $n+1$ transition depends on the value of process at the previous $n$-th transition and the value of the index process at the previous $n$-th transition. The index process has been introduced to incorporate past information that contributes to the composition of next return. The use of a moving average is done in order to exclude remote information that are not determinant in the formation of the new return. \\
\indent Equation \eqref{hp} asserts that for each value of the index process $v\in \mathbb{R}$ there is a matrix ${\bf{P}}(v)=(p_{ij}(v))_{i,j\in E}$ that gives the probability of transitions among the states. In order to apply the model and to reduce unnecessary parameterizations it is necessary to identify for which values of the index process an actual change of dynamics has to be considered, in other words it is necessary to discretize the values of the index process in a given number of states where an effective change of dynamic occurs.


\section{Discretization of the state space through the change point approach}
\label{sec:chp}

In this section, we develop the idea of adopting the change-point approach for Markov chain as presented by \cite{polansky2007detecting} to our indexed model with the notable difference that our change points are not times but values of the index process in correspondence of which a change of dynamic occur. Hence, our target is to determine the optimal number of states of the index process $V_{n}^{m}$ as well as the border values for each state. This will be obtained by using the fact that the dynamics of the price return process $J_n$ depend on the value of the index process $V_{n}^{m}$.  

Let us assume that we have fixed a value of the memory $m$ and that we have observed a trajectory of the price process $S_{n}$ from time zero to $T\in \mathbb{N}$ that is the time of the last observation of the process. Then, from the observed time series of log-returns $\{J_{n}\}_{n=0}^{T}$ we can construct, using relation $(\ref{index})$, the corresponding time series of the index process $\{V_{n}^{m}\}_{n=0}^{T}$.  Since the function $f$ is bounded and the set $E$ is finite, the process  $V_{n}^{m}$ assumes value between a maximum value $\overline{E}_{f}$ and a minimum value $\underline{E}_{f}$, then: 
\begin{equation}
V_{n}^{m}(\omega) \in [\underline{E}_{f},\overline{E}_{f}] .
\end{equation}

First, let us suppose that there are mainly two levels of volatility in the market: low volatility and high volatility. This is equivalent to find a value of the index process that represents a change-point in the return dynamic. What we are interested in is to model the price return process $J_n$ through two different Markov processes described by two different transition probability matrices. Let $\underline{P}(\psi_{1})$ be the transition probability matrix of the price return process in case of low volatility and $\overline{P}(\psi_{1})$ be the transition probability matrix in case of high volatility. 

Since the level of the volatility is considered through the index process then, the state space $[\underline{E}_{f},\overline{E}_{f}]$ of the Index process has to be subdivided into two discrete states. Let $\psi_{1} \in [\underline{E}_{f},\overline{E}_{f}] $ be the value of the Index process that determines a change in the dynamics of the price return process $J_n$ such that:

\begin{enumerate}[(a)]
	\item The interval $[\underline{E}_f,\psi_{1})$ represents the low volatility case. If the Index process $V_{n}^{m}<\psi_{1} $, then we suppose that the price return process $J_n$ is described by the transition probability matrix $\underline{P}(\psi_{1})$;
	\item The interval $[\psi_{1},\overline{E}_{f}]$ represents the high volatility case. If the Index process $V_{n}^{m}\geq\psi_{1} $, then we suppose that the price return process $J_n$ is described by the transition probability matrix $\overline{P}(\psi_{1})$.
\end{enumerate}  

\paragraph{Case of one known change point}

If the change point $\psi_{1}$ is known, the transition probability matrices can be estimated through the maximum likelihood estimators $\hat{\overline{P}}$ and $\hat{\underline{P}}$ whose $(i,j)$-th elements are given as follows:
\begin{equation}
\hat{\overline{P}}_{ij}(\psi_{1})=\frac{\sum_{n=1}^{T} \mathbf{1}_{\{J_{n-1}=i,J_{n}=j,V_{n-1}^{m}\geq\psi_{1}\}}}
{\sum_{n=1}^{T}\mathbf{1}_{\{J_{n-1}=i,V_{n-1}^{m}\geq\psi_{1}\}}}=:\frac{\overline{N}_{ij}(\psi_{1})}{\overline{N}_{i}(\psi_{1})} 
\label{overlineP}
\end{equation}						
\smallskip
\begin{equation}
\hat{\underline{P}}_{ij}(\psi_{1})=\frac{\sum_{n=1}^{T} \mathbf{1}_{\{J_{n-1}=i,J_{n}=j,V_{n-1}^{m}<\psi_{1}\}}}
{\sum_{n=1}^{T}\mathbf{1}_{\{J_{n-1}=i,V_{n-1}^{m}<\psi_{1}\}}}=:\frac{\underline{N}_{ij}(\psi_{1})}{\underline{N}_{i}(\psi_{1})} .
\label{underlineP}
\end{equation}

In order to test the hypothesis that the two matrices are statistically equal and, as a consequence, there is no significant change in the dynamics of the price return process due to the different levels of the states of the index process, a statistical test can be developed at a given significance level $\alpha$:	
\begin{equation}
\left\{ \begin{aligned}
&H_{0}: \overline{P}=\underline{P}\,, \\
&H_{1}: \overline{P}\neq \underline{P}\,.
\end{aligned} \right. 
\label{test}
\end{equation}	

Once the two hypotheses are set, a distance measure has to be defined. As shown in \cite{polansky2007detecting}, a convenient distance measure can be developed through the log of the likelihood ratio. 
Let $\mathfrak{L}(\overline{P}(\psi_{1}),\underline{P}(\psi_{1}),\underline{\mathit{x}})$ be the likelihood function of a sample $\underline{x}$ observed on the period $[0,T]$ with a change point $\psi_{1}$. Then,
\begin{equation}
\begin{aligned}
&\mathfrak{L}(\overline{P}(\psi_{1}),\underline{P}(\psi_{1}),\underline{\mathit{x}})= \mathbb{P}(J_{0}=x_{0},J_{1}=x_{1},\dots,J_{T}=x_{T}) \\
&=\prod_{i,j \in E} (\overline{P}_{ij}(\psi_{1}))^{\overline{N}_{ij}(\psi_{1})}*\prod_{i,j \in E} (\underline{P}_{ij}(\psi_{1}))^{\underline{N}_{ij}(\psi_{1})}.
\end{aligned}
\end{equation}

Let $\mathit{L}$ be the log-likelihood function:						
\begin{equation}
\begin{aligned}[left]
&\mathit{L}(\overline{P}(\psi_{1}),\underline{P}(\psi_{1}),\mathit{\underline{x}})=log(\mathit{L}) \\
&=\sum_{ij}\bigg(\overline{N}_{i,j}(\psi_{1})*
log\bigg(\frac{\overline{N}_{i,j}(\psi_{1})}{\overline{N}_{i}(\psi_{1})}\bigg)\bigg) +
\sum_{ij}\bigg(\underline{N}_{i,j}(\psi_{1})*
log\bigg(\frac{\underline{N}_{i,j}(\psi_{1})}{\underline{N}_{i}(\psi_{1})}\bigg)\bigg)\\
&=:L(\overline{P}(\psi_{1}),\underline{x})+L(\underline{P}(\psi_{1}),\underline{x}).
\label{loglik}
\end{aligned}
\end{equation}

A distance measure can then be calculated from the log-likelihood functions:
\begin{equation}
\mathfrak{D}= 2*[L(\overline{P}(\psi_{1}),\underline{x})+L(\underline{P}(\psi_{1}),\underline{x})-L(P,\underline{x})],		
\label{D}
\end{equation}
where
\begin{itemize}
	\item $L(\overline{P}(\psi_{1}),\underline{x})+L(\underline{P}(\psi_{1}),\underline{x})$ is the log of the likelihood value for the model under the alternative hypotheses;
	\item $L(P,\underline{x})$ is the log of the likelihood value for model under the null hypotheses. In this case, we suppose that there is no change point in the observed process and thus we estimate the dynamics of the price return process $\{J_n\}$ by using a single transition probability matrix $P$ of the Markov Chain. 
\end{itemize}

Standard asymptotic theory (see, e.g. \cite{bill61}) can be used to show that if $T\rightarrow \infty$ and $\overline{N}_{i,j}(\psi_{1})\rightarrow \infty$ and $\underline{N}_{i,j}(\psi_{1})\rightarrow \infty$ for all states $i\in E$, then $\mathfrak{D}$ converges in distribution to a $\chi$-squared random variable with $\mid E\mid \cdot (\mid E\mid -1)$ degrees of freedom.

\paragraph{Case of one unknown change points}

In our case, the change point is not known and it also needs to be estimated. In cases in which the change point is not known, the parameter $\psi_{1}$ becomes an unknown parameter of the log-likelihood function seen above. Usually, the maximum likelihood estimator for $\psi_{1}$, in this cases, does not exist in closed form but it can be estimated through an iterative method. Operatively, the methodology is the following:

\begin{enumerate}
	\item Define on the state space $n$ discrete and distinct points $\psi\in [\underline{E}_f,\overline{E}_f]$;
	\item For each possible ${\psi}\in [\underline{E}_f,\overline{E}_f]$
	
	\begin{enumerate}
		\item Set $\psi_1=\psi$;
		\item Estimate $\overline{P}(\psi_1)$ with Eq. \eqref{overlineP} and $\underline{P}(\psi_1)$ with Eq. \eqref{underlineP};
		\item Compute $L(\overline{P}(\psi_1),\underline{P}(\psi_{1}),\underline{x})=L(\overline{P}(\psi_1),\underline{x})+L(\underline{P}(\psi_1),\underline{x})$ as in Eq. \eqref{loglik};
	\end{enumerate}
	\item Fix $\hat{\psi}_{1}= arg \  max\{\psi_1\in [\underline{E}_f,\overline{E}_f]: L(\overline{P}(\psi_1),\underline{P}(\psi_{1}),\underline{x}) \}$.
\end{enumerate}	

The result will be the value of the index process that determines a change in the price return process, the estimated transition probability matrices of the price return process that reflect the two different dynamics and the value of the log-likelihood function. Once these outputs are obtained, a test of hypothesis as shown in \eqref{test} can be performed. 

Unfortunately, the theoretical distribution of the statistic test under $H_0$ is also not known but it can be approximated by using the bootstrap methodology as for the detection of a time changing point for Markov chains, see \cite{polansky2007detecting}. Given a sample $\underline{x}$, a single transition probability matrix $P$ is estimated from the data, i.e. without considering the change point. A number $B$ of trajectories of the same length of the data are simulated from the transition matrix $P$. For each simulation, the test statistic defined in Eq. \eqref{D} is calculated. We denote this values as $D_B$. The theoretical distribution of the statistic test $D$ can then be approximated by the kernel distribution of the simulated statistic test $D_B$. 

Once the level of confidence $\alpha$ for the test is fixed, the critical value $d_\alpha$ can be approximated by the $1-\alpha$ percentile of the simulated statistic test $D_B$. Hence, if the statistic test on the sample data $\hat{D}(\underline{x}) \geq d_{\alpha}$, the null hypothesis will be refused.     

The p-value of the test can then be  calculated as:
\begin{equation}
p-value=\frac{1}{B+1}*\Big[1+\sum_{i=1}^{B}\mathbf{1}_{\{D_B(\underline{x})\geq \hat{D}(\underline{x})\}}\Big]
\label{pvalue}
\end{equation}			

\paragraph{Case of more than one unknown change points} 
The case of two or more change points can be extended from the estimation methods of a single change point shown so far.

Let $\psi_{1}<\psi_{2}<\dots<\psi_{k}$ be $k$ change points defined in the interval $[\underline{E}_{f},\overline{E}_{f}]	$. Each possible combination $(s)$ of the $k$ change points divides the interval in $k+1$ sub-intervals:
\[	\begin{cases}
[\underline{E}_{f},\psi_{1}] & =:I_{1}^{(s)}\\
(\psi_{1},\psi_{2}] & =:I_{2}^{(s)}\\
\quad \,\, \vdots \\
(\psi_{r-1},\psi_{r}] & =:I_{r}^{(s)}\\
\quad \,\, \vdots\\
(\psi_{k},\overline{E}_{f}] & =:I_{k+1}^{(s)} .
\end{cases}
\] 

For each of the $(s)$ possible partitions of the interval in $k+1$ sub-intervals $\big\{I_{r}^{(s)}\big\}_{r=1}^{k+1}$, $k+1$ transition probability matrices $\hat{P}^{(s)}$ have to be estimated whose $(i,j)$ elements are:
\begin{equation}
\hat{P}_{i,j;r}^{(s)}=\frac{\sum_{n=1}^{T}\mathbf{1}_{\{J_{n-1}=i,J_{n}=j,V_{n-1}^{m}\in I_{r}^{(s)}\}}}
{\sum_{n=1}^{T}\mathbf{1}_{\{J_{n-1}=i,V_{n-1}^{m}\in I_{r}^{(s)}\}}}=:\frac{N_{i,j;r}^{(s)}}{N_{i;r}^{(s)}}	.		
\end{equation}

The likelihood function of the sample $\underline{\mathit{x}}$ observed on the period $[0,T]$ for the generic partition $§(s)$ with $k+1$ change points is:
\begin{equation}
\mathfrak{L}^{(s)}(P_{1}^{(s)},P_{2}^{(s)},\dots,P_{k+1}^{(s)};\underline{\mathit{x}})=
\prod_{r=1}^{k+1}\prod_{i,j \in E}\big(P_{i,j;r}^{(s)}\big)^{N_{i,j;r}^{(s)}}.
\end{equation}

The log-likelihood function can then be calculated as
\begin{equation}
L^{(s)}=log(\mathfrak{L}^{(s)})=L_{1}^{(s)}+L_{2}^{(s)}+
\dots+L_{k+1}^{(s)}=\sum_{r=1}^{k+1}L_{r}^{(s)},
\end{equation}
where $L_{r}^{(s)}:=L(P_{r}^{s},\underline{x})$ is the log-likelihood of the generic $(s)$ partition in reference to the sub-interval $I_r$.
As seen above, the values of the $k$ change points can be calculated in an iterative way. We set $n$ discrete and distinct values of $\psi\in [\underline{E}_f,\overline{E}_f]$. Notice that, for $k$ change points, we have $\binom{n-1}{k}$ possible combinations of the vector of change points $(\psi_1,\psi_2,\dots,\psi_k)$. The methodology is basically the same as that for one change point but in this case we have to consider all the possible combinations of the set of $k$ change points $(\psi_1,\psi_2,\dots,\psi_k)$. The estimated values can be calculated through:
\begin{equation}
(\hat{\psi}_1,\hat{\psi}_2,\dots,\hat{\psi}_k)=arg \  max\{(\psi_1<\psi_2<\dots<\psi_k)\in [\underline{E}_f,\overline{E}_f]:L^{(s)}\}
\end{equation}

A test can be performed in this case where:		
\begin{equation}
\left\{ \begin{aligned}
&H_{0}: P_0=P_1=\dots=P_k \\
&H_{1}:  P_i\neq P_j & \text{for some} \ i\neq j
\end{aligned} \right. 
\label{testmulti}
\end{equation}	

The test statistic can be calculated as follows:
\begin{equation}
\mathfrak{D}^{(s)}=2*\Big[\sum_{r=1}^{k+1}L_{r}^{(s)}-L(P)\Big]
\end{equation}
where $L(P)$ is the log-likelihood value in case there is no change point on the sample $\underline{x}$ and thus a single transition matrix $P$ is estimated. The bootstrap methodology can be used again to obtain an approximation of the theoretical distribution of the statistic test.  

If the number of change points is unknown, it also needs to be estimated and thus measures such as AIC or BIC can be used. 

The AIC objective function is given by:
\begin{equation}
AIC(k)=2*\mid E\mid*(\mid E\mid-1)*(k+1)-2*\sum_{r=1}^{k+1}L_{r}^M
\end{equation}	
where
\begin{itemize}
	\item $L^M$ is the value of the log-likelihood of the maximum likelihood estimate of ${\psi_1,\psi_2,\dots,\psi_k}$ conditioned on the fact that there are $k$ change points
	\item $\mid E\mid$ is the number of states of the Markov Chain
	\item $k$ is the number of change points.
\end{itemize}	
The estimated number of change points is given by:
\begin{equation}
\label{AIC}
\hat{k}_{AIC}=arg \ min \{k \in \{1,2,\dots,n\}:AIC(k)\}.
\end{equation}

Alternatively one can use the BIC criterion given by:
\begin{equation}
BIC(k)=2*log(n)*\mid E\mid*(\mid E\mid-1)*(k+1)-2*\sum_{r=1}^{k+1}L_{r}^{M} .
\end{equation}	
The estimated number of change points is given by:
\begin{equation}
\label{BIC}
\hat{k}_{BIC}=arg \ min \{k \in \{1,2,\dots,n\}:BIC(k)\}.
\end{equation}

\section{Predicting the value of the index in the next step}
\label{sec:predict}

Suppose we have identified $k$ change points of our Indexed Markov chain $\{J_n\}_{n\in\mathbb{N}}$ such that:
\begin{equation*}
\psi_1 < \psi_2 < \dots < \psi_k \ ; \ \psi_i\in\mathbb{R} \,\ \forall i = 1,2, \dots, k.
\end{equation*}

Accordingly we have estimated $k+1$ transition probability matrices denoted by:
\begin{equation*}
P(v)=(p_{ij}(v))_{i,j\in E,v\in \mathbb{R}}
\end{equation*}
where $p_{ij}(u)=p_{ij}(v)$ if $\: \exists a \in \{1,2, \dots, k-1\}$ such that
\[\: \psi_a < u \leq \psi_{a+1} \ ; \ \psi_a<v\leq\psi_{a+1}.\]

This means that if $u,v$ belong to the same interval $I_{a+1}:=(\psi_a,\psi_{a+1}]$ where $I_1=(-\infty, \psi_1]$ and $I_{k+1}=(\psi_{k}, +\infty]$, then $P(u)=P(v)$.
 
At time $s$, the information needed in order to apply the model is the vector of the past states of the return process ${\bf{i}}_{s-m+1}^s:=\{i_{s-m+1},i_{s-m+2}, \dots, i_{s-1}, i_{s}\}$. Once the vector ${\bf{i}}_{s-m+1}^s$ is known, we can calculate the value of the index process: 
\begin{equation}
V_s^m=\frac{\sum_{k=0}^{m-1}f(J_{s-k})}{m}=\frac{\sum_{k=0}^{m-1}f(i_{s-k})}{m}.
\end{equation}

Let us define:
\begin{equation}
T_{{\bf{i}}_{s-m+1}^s}(I_a):= \inf \{ n\geq s, n\in \mathbb{N}: V_n^m\in I_a | {\bf{J}}_{s-m+1}^s={\bf{i}}_{s-m+1}^s\}, 
\end{equation}
as the first time (successive of the current time $s$) in which the Index process enters into the generic interval $I_a=(\psi_{a-1},\psi_a]$. 

Also, let
\begin{equation}
g_{{\bf{i}}_{s-m+1}^s}(I_a;s+n)=\mathbb{P}(T_{{\bf{i}}_{s-m+1}^s}(I_a)=s+n|{\bf{J}}_{s-m+1}^{s}={\bf{i}}_{s-m+1}^s),
\end{equation}
be the probability distribution of the first entrance time in $I_a$ of the Index process. Let us assume that
\[g_{{\bf{i}}_{s-m+1}^s}(I_a;s)=0 \ ; \ \forall I_a \ , \ \forall {\bf{i}}_{s-m+1}^s,\] then the following result holds true:

\begin{Proposition}
	\label{prep1}
	(explicit formula for $g$)
	\begin{equation}
\label{g}
	g_{{\bf{i}}_{s-m+1}^s}(I_a;s+n)=\!\!\!\!\!\sum_{(i_{s+1},\dots,i_{s+m-1}),i_{s+n}}^{E^c,E} \prod_{r=1}^{n} P_{i_{s+r-1},i_{s+r}} \bigg(\frac{\sum_{k=0}^{m-1}f(i_{s+r-1-k})}{m}\bigg),
	\end{equation}
\end{Proposition}
where the symbol
\begin{multline*}
\sum_{(i_{s+1},\dots,i_{s+m-1}),i_{s+n}}^{E^c,E} := \sum_{i_{s+1}\in E^c_{{\bf{i}}_{s-m+1}^0}(I_a)} \: \ \sum_{i_{s+2}\in E^c_{{\bf{i}}_{s-m+2}^1}(I_a)} \dots \\ \dots \sum_{i_{s+m-1}\in E^c_{{\bf{i}}_{s-m+(n-1)}^{s+n-2}}(I_a)} \:\ \sum_{i_{s+n}\in E^c_{{\bf{i}}_{s-m+n}^{s+n-1}}(I_a)} 
\end{multline*} 
and
\[E_{{\bf{i}}_{s-m+1}^s}(I_a):=\{k \in E: \frac{\sum_{k=0}^{m-1} f(i_{s+1-k})}{m} \in I_a \},\]
also, 
$E^c_{{\bf{i}}_{s-m+1}^s}(I_a)$ is the complementary of $E_{{\bf{i}}_{s-m+1}^s}(I_a)$.

\begin{proof}
First of all it should be remarked that $E_{{\bf{i}}_{s-m+1}^s}(I_a)$ is the sub-interval of the space state $E$ that, through the estimation of ${\bf{i}}^s_{s-m+1}$, let the Index process enter in the interval $I_a$ with the next transition. Next, we can start to prove the result by using mathematical induction.

	For $n=1$, Proposition \ref{prep1} is true. In fact:
	\begin{equation}
	\begin{aligned}
	g&_{{\bf{i}}_{s-m+1}^s}(I_a;s+1)=\mathbb{P}(T_{{\bf{i}}_{s-m+1}^s}(I_a)=s+1|{\bf{J}}_{s-m+1}^s={\bf{i}}_{s-m+1}^s) \\
	&=\sum_{i_{s+1}\in E} \mathbb{P}(T_{{\bf{i}}_{s-m+1}^s}(I_a)=s+1,J(s+1)=i_{s+1}|{\bf{J}}^s_{s-m+1}={\bf{i}}^s_{s-m+1})\\
	&=\sum_{i_{s+1}\in E} \mathbb{P}(V_{s+1}^m \in (I_a), J(s+1)=i_{s+1}|{\bf{J}}^s_{s-m+1}={\bf{i}}^s_{s-m+1})\\
	&=\sum_{i_{s+1}\in E} \mathbb{P}(V_{s+1}^m \in (I_a)| {\bf{J}}^{s+1}_{s-m+1}={\bf{i}}^{s+1}_{s-m+1})\\
	&*\mathbb{P}(J(s+1)=i_{s+1}| {\bf{J}}^{s}_{s-m+1}={\bf{i}}^{s}_{s-m+1}))\\
	&=\sum_{i_{s+1}\in E_{{\bf{i}}^s_{s-m+1}}(I_a)} 1* {P}_{i_s,i_{s+1}}\bigg(\frac{\sum_{k=0}^{m-1}f(i_{s-k})}{m}\bigg),
	\end{aligned}
	\end{equation}
that is equal to formula $(\ref{g})$ for $n=1$.

Now, let us suppose that Preposition \ref{prep1} is true for $n-1$, which means that:
\begin{equation*}
g_{{\bf{i}}_{s-m+1}^s}(I_a;s+m-1)=\sum_{(i_{s+1},\dots,i_{s+m-2},i_{s+m-1})}^{E^c,E} \prod_{r=1}^{(s+m-1)-s}{P}_{i_{s+r-1},i_{s+r}}\bigg(\frac{\sum_{k=0}^{m-1}f(i_{s-k})}{m}\bigg).
\end{equation*} 
We can calculate:
\begin{equation}
\begin{aligned}
g_{{\bf{i}}_{s-m+1}^s}(I_a;s+n)&=\mathbb{P}(T_{{\bf{i}}_{s-m+1}^s}(I_a)=s+n|{\bf{J}}_{s-m+1}^s={\bf{i}}_{s-m+1}^s)\\
=\sum_{i_{s+1}\in{E_{{\bf{i}}_{s-m+1}^s}^c(I_a)}}&\mathbb{P}(T_{{\bf{i}}_{s-m+1}^s}(I_a)=s+n,J(s+1)=i_{s+1}|{\bf{J}}_{s-m+1}^s={\bf{i}}_{s-m+1}^s)\\
=\sum_{i_{s+1}\in{E_{{\bf{i}}_{s-m+1}^s}^c(I_a)}}&\mathbb{P}(T_{{\bf{i}}_{s-m+1}^s}(I_a)=s+n|J(s+1)=i_{s+1},{\bf{J}}_{s-m+1}^s={\bf{i}}_{s-m+1}^s)\\
&*\mathbb{P}(J(s+1)=i_{s+1}|{\bf{J}}_{s-m+1}^s={\bf{i}}_{s-m+1}^s)\\
\end{aligned}
\end{equation}
\begin{equation}
\begin{aligned}
&=\sum_{i_{s+1}\in{E_{{\bf{i}}_{s-m+1}^s}^c(I_a)}} \mathbb{P}(T_{{\bf{i}}_{s-m+1}^s}(I_a)=s+n|{\bf{J}}_{s-m+1}^s={\bf{i}}_{s-m+1}^s)\\
& * {P}_{i_s,i_{s+1}}\bigg(\frac{\sum_{k=0}^{m-1}f(i_{s-k})}{m}\bigg)\\
& =\sum_{i_{s+1}\in{E_{{\bf{i}}_{s-m+1}^s}^c(I_a)}}g_{{\bf{i}}_{s-m+2}^{s+1}}(I_a;s+n)* {P}_{i_s,i_{s+1}}\bigg(\frac{\sum_{k=0}^{m-1}f(i_{s-k})}{m}\bigg)\\
& = (\text{from \ inductive \ hypothesis})\\
& =\sum_{i_{s+1}\in{E_{{\bf{i}}_{s-m+1}^s}^c(I_a)}}\bigg[\sum_{(i_{s+2},\dots,i_{s+m-2}),i_{s+m-1}}^{E^c,E}\\
& \prod_{r=1}^{s+n-(s+1)} {P}_{i_{s+1-r-1},i_{s+1+r}}\bigg(\frac{\sum_{k=0}^{m-1}f(i_{s+1,r-1-k})}{m}\bigg)\bigg]* {P}_{i_s,i_{s+1}}\bigg(\frac{\sum_{k=0}^{m-1}f(i_{s-k})}{m}\bigg),
\end{aligned}
\end{equation}
where the latter coincides with formula $(\ref{g})$.
\end{proof}

\section{Empirical study}
\label{sec:emp}
The methodology described so far has been applied to the study of intra-day prices of ENI, a quoted Italian firm. 
The sample data of prices starts on January $1^{st}$, 2007 and ends on December $31^{st}$ 2010. 
The dataset is obtained from $www.borsaitaliana.it$ and it contains tick-by-tick quotes of the traded stock. The data have been re-sampled to have 1 minute frequency.
The log returns are calculated from the value of the financial asset under study $S_{n}$: $R_{n}=\log(S_{n}/S_{n-1})$ and then discretized through map given by Eq. $(\ref{map})$. According to \cite{d2011semi} returns have been discretized into 5 states chosen to be symmetrical with respect to returns equal zero and to keep the shape of the distribution unchanged. Following the results obtained from the same authors in \cite{d2011semi} $f$, defined in Eq. \ref{index}, is chosen to be simply equal to $J^2$. The value of $m$ is fixed to be equal to $30$ minutes. Then the index process $V^m_n$ is given by the moving average up to past $30$ minutes of the square of returns.   The value of the index denotes the level of the volatility of the prices:  higher the value of the index, higher the price volatility. 

Firstly, let us suppose that there can be only two levels of the volatility: high and low. As a consequence, the index process should also be discretized into these two states. In order to detect the optimal value of the index that subdivides the process in two states we use the fact that the price return process $\{J_n\}$ presents different dynamics, i.e. different transition matrices, based on the level of the volatility. Thus, in our work the change point is identified as the value of the index which would maximize the differences in the price return dynamics. 

\begin{figure}
	\centering
	\includegraphics[width=0.8\textwidth]{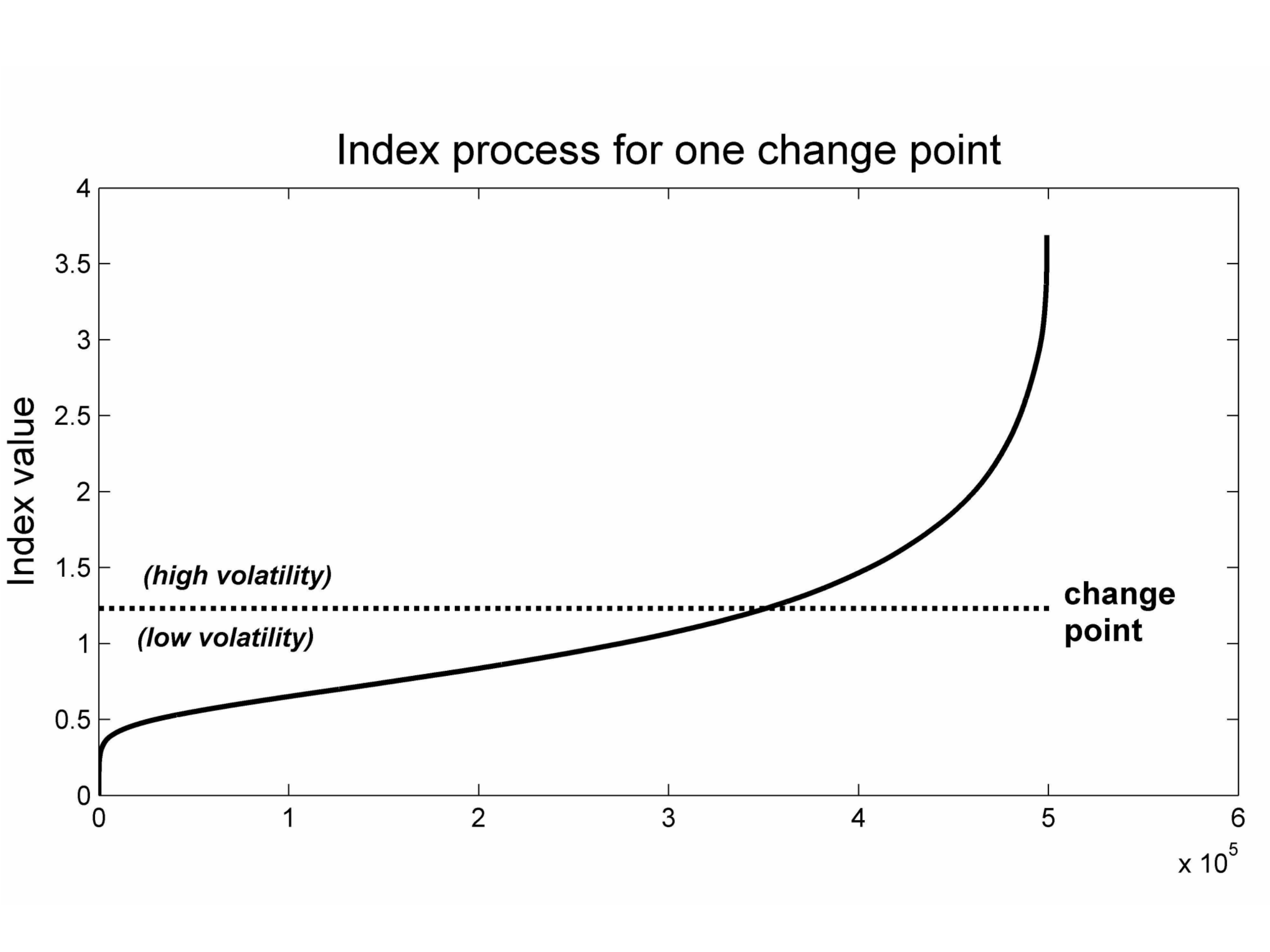}
	\caption{Discretization of the index process in case of one change point }
	\label{fig:index}	
\end{figure}

Since there is no closed form for identifying the change point we operate algorithmically by using the maximum likelihood procedure described in Section \ref{sec:chp}. The estimated change point and the two probability matrices of the price return process in each state of the index are calculated. Figure \ref{fig:index} shows the discretization of the index process in two states by using the change point identified through the maximum log-likelihood procedure. 

A well known characteristic of financial time series is the "volatility clustering" which simply states that periods of high/low volatility in the market tend to be followed by periods of high/low volatility. It seems useful to point out that this characteristic is confirmed by analyzing the following two transition probability matrices: 

\tiny
\[\hat{\overline{P}}(\psi_1)=
\begin{pmatrix}
0.173 & 0.151 & 0.207 & 0.218 & 0.251 \\
0.129 & 0.196 & 0.267 & 0.255 & 0.153 \\
0.137 & 0.209 & 0.300 & 0.221 & 0.133 \\
0.150 & 0.246 & 0.275 & 0.204 & 0.125 \\
0.237 & 0.215 & 0.218 & 0.159 & 0.171
\end{pmatrix} \ ; \
\hat{\underline{P}}(\psi_1)=
\begin{pmatrix}
0.067 & 0.162 & 0.312 & 0.338 & 0.121 \\
0.031 & 0.183 & 0.391 & 0.347 & 0.048 \\
0.033 & 0.236 & 0.466 & 0.234 & 0.031 \\
0.049 & 0.339 & 0.397 & 0.185 & 0.030 \\
0.110 & 0.338 & 0.316 & 0.170 & 0.066
\end{pmatrix}
\]

\normalsize

In fact, it can be noticed that in the case of high volatility the transition matrix $\hat{\overline{P}}(\psi_1)$ presents higher probabilities for the most extreme states than the transition matrix $\hat{\underline{P}}(\psi_1)$. This simply confirms the fact that in a high volatility market, the probability of experiencing large variations of the asset price (i.e. strongly positive returns followed by strongly negative returns and vice versa) is higher than the case of low volatility.

Once the matrices are estimated, the next step is that of determining whether they are statistically different. For this purpose we calculated two indices that can be used as a measure of the difference between the two matrices. In particular, we used the \emph{percentage root mean square deviation (\% RSMD)} and the \emph{percentage mean absolute deviation (\%MAD)}. The formulas are given in Eq. \ref{eq:rsmd} and Eq. \ref{eq:mad}, respectively. 

\begin{equation}
\label{eq:rsmd}
\% RSMD = \sqrt{\frac{\sum_{ij}(\overline{p}_{ij} - \underline{p}_{ij})^2}{n}}*\frac{n * 100\%}{\sum_{ij}\underline{p}_{ij}},
\end{equation}

\begin{equation}
\label{eq:mad}
\% MAD = \frac{\sum_{ij}|\overline{p}_{ij} - \underline{p}_{ij}|}{\sum_{ij}\underline{p}_{ij}}*100\% ,
\end{equation}

Results are shown in Table \ref{rmsd}. Since, the values of these two indices are a measure of the differences of the estimated matrices, higher the value of the index, higher the distance between the two matrices and higher the probability that there actually is a  difference in the process.

\begin{table}
	\begin{center}
		\small
		\caption{\% Root square mean deviation and \% mean absolute deviation} 	
		\begin{tabular}{c|cc}
			\label{rmsd}	
			Matrices & \% RSMD & \% MAD  \\ \hline
			$\hat{\overline{P}}(\psi_1)$ , $\hat{\underline{P}}(\psi_1)$ & 49.3\% & 44.2\% \\ 
		\end{tabular} 	
	\end{center}
\end{table} 

The next step is to test whether the two transition probability matrices are statistically different. In order to construct the test, we use the procedure described in Section \ref{sec:chp}.    

\begin{figure}
	\centering
	\includegraphics[width=0.8\textwidth]{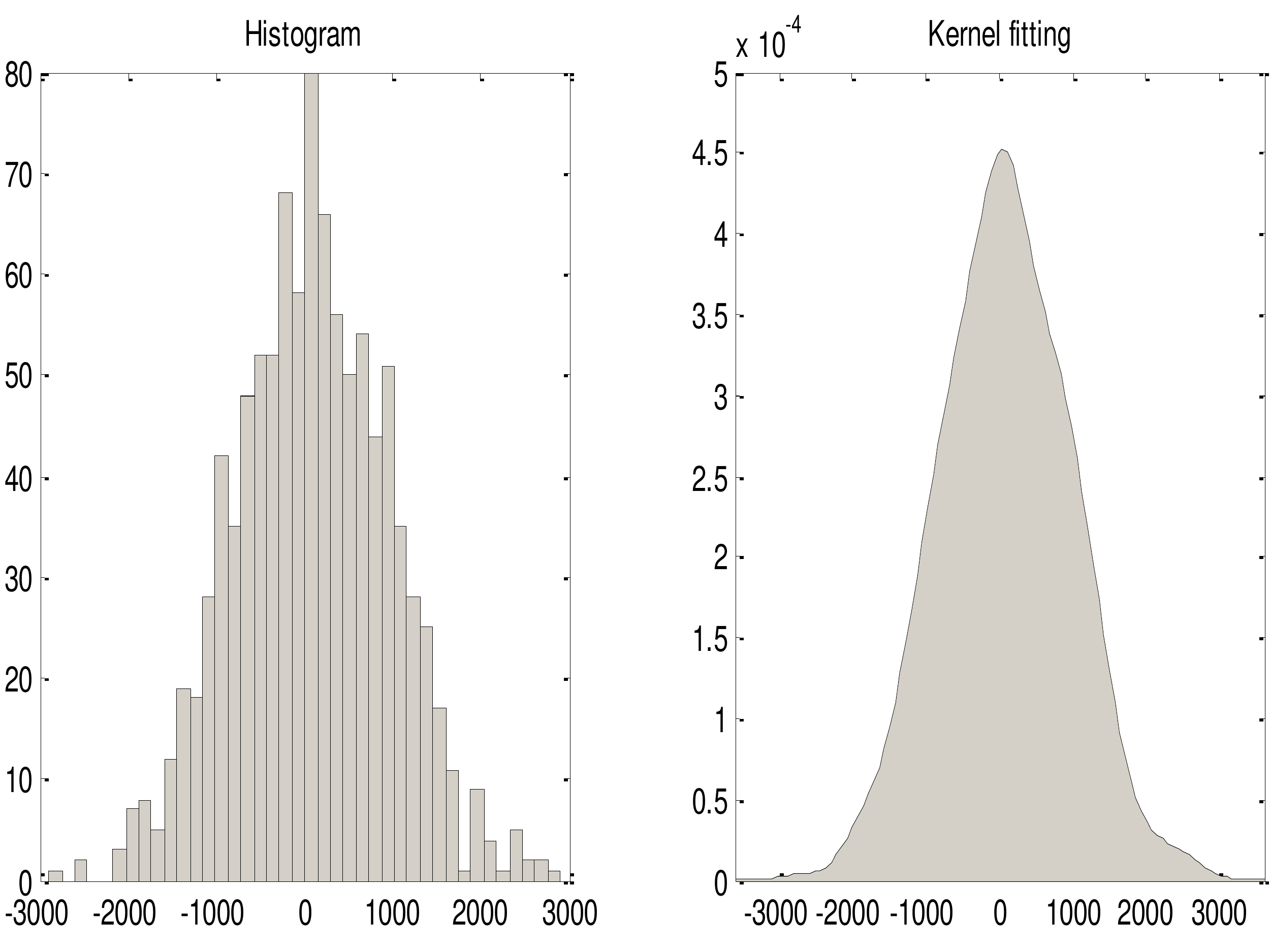}
	\caption{Histogram and kernel fitting of the empirical statistic test}
	\label{fig:hist}	
\end{figure}

As mentioned before, the distribution of the statistic test under the null hypothesis is not known and thus it has to be approximated through the bootstrap methodology. We have simulated \emph{1000} trajectories of the same length of the data from a single transition probability matrix estimated considering the whole dataset. The histogram of the simulated statistic test as well as the kernel fitting are derived and are shown in Figure \ref{fig:hist}.
\begin{table}
	\begin{center}
		\small
		\caption{Results of the statistic test on one change point} 	
		\begin{tabular}{c|ccccc}
			\label{res1chp}	
			& $\psi_1$ & $\mathbf{\mathfrak{D}}$ & $\mathbf{\mathfrak{D^*(0.95)}}$ & $\mathbf{\mathfrak{D^*(0.99)}}$ & Empirical p-value  \\ \hline
		Value & 1.23 & 32400 & 3290 & 3580 & 0.000 \\  
		\end{tabular} 	
	\end{center}
\end{table} 

We chose a level of significance of $5\%$. The results of the test for a single change point of the stock are shown in Table \ref{res1chp}. $\mathfrak{D^*(0.95)}$ and $\mathfrak{D^*(0.99)}$ are derived from the simulated distribution and represent the value of the statistic test at a 95\% and a 99\% confidence level, respectively. Since there does not exist a theoretical distribution of the statistic test and, as a consequence, it does not exist a theoretical p-value, we obtained the p-value from the simulated distribution as well, and reported it as the 'empirical' p-value. The results of the statistic test strongly suggest to not accept the null hypothesis and thus we conclude that there is a statistical difference between the two transition probability matrices. 

%
%
%

The case of more then one change point has also been considered. The procedure is the same as for the case of one change point described so far. The main difference is that in this case, methods based on AIC and BIC as defined in Eq. \eqref{AIC} and Eq. \eqref{BIC}, respectively, can be used to identify the most parsimonious model. As mentioned, it does not exist a closed form for the calculation of the optimal number of change points and thus we operate in an iterative way.

It should be remarked that the algorithm reaches very slowly the minimum values of AIC or BIC. We decided to also calculated the improvement of an additional change point in terms of a percentage variation of the two indices. We considered a limit of less than 0.1\% as a sign that the algorithm is reaching its minimum. Results are summarized in Table \ref{resmultchp}.   

\begin{table}
\begin{center}
	\caption{Results of the statistic test on various number of change points} 
	\begin{tabular}{c|ccccc}
	\label{resmultchp}	
		$\textbf{k}$ & \textbf{1} & \textbf{2} & \textbf{3} & \textbf{4} & \textbf{5} \\ \hline
		$\mathbf{\mathfrak{D}}$ & 32400 & 42000 & 46100 & 48300 & 49800 \\
		\textit{$\% \Delta$} & & 29.6\% & 9.8\% & 4.8\% & 3.1\% \\ \hline
		\textbf{AIC} & 1379000 & 1370000 & 1365000 & 1363000  & 1362000 \\
		\textit{$\% \Delta$} & & -0.7\% & -0.4\% & -0.1\% & $<$-0.1\% \\ \hline
		\textbf{BIC} & 1380000 & 1371000 & 1367000 & 1366000 & 1365000 \\
		\textit{$\% \Delta$} & & -0.7\% & -0.3\% & $<$-0.1\% & $<$-0.1\% \\
	\end{tabular} 
\end{center}
\end{table}

\begin{figure}
	\centering
	\includegraphics[width=0.8\textwidth]{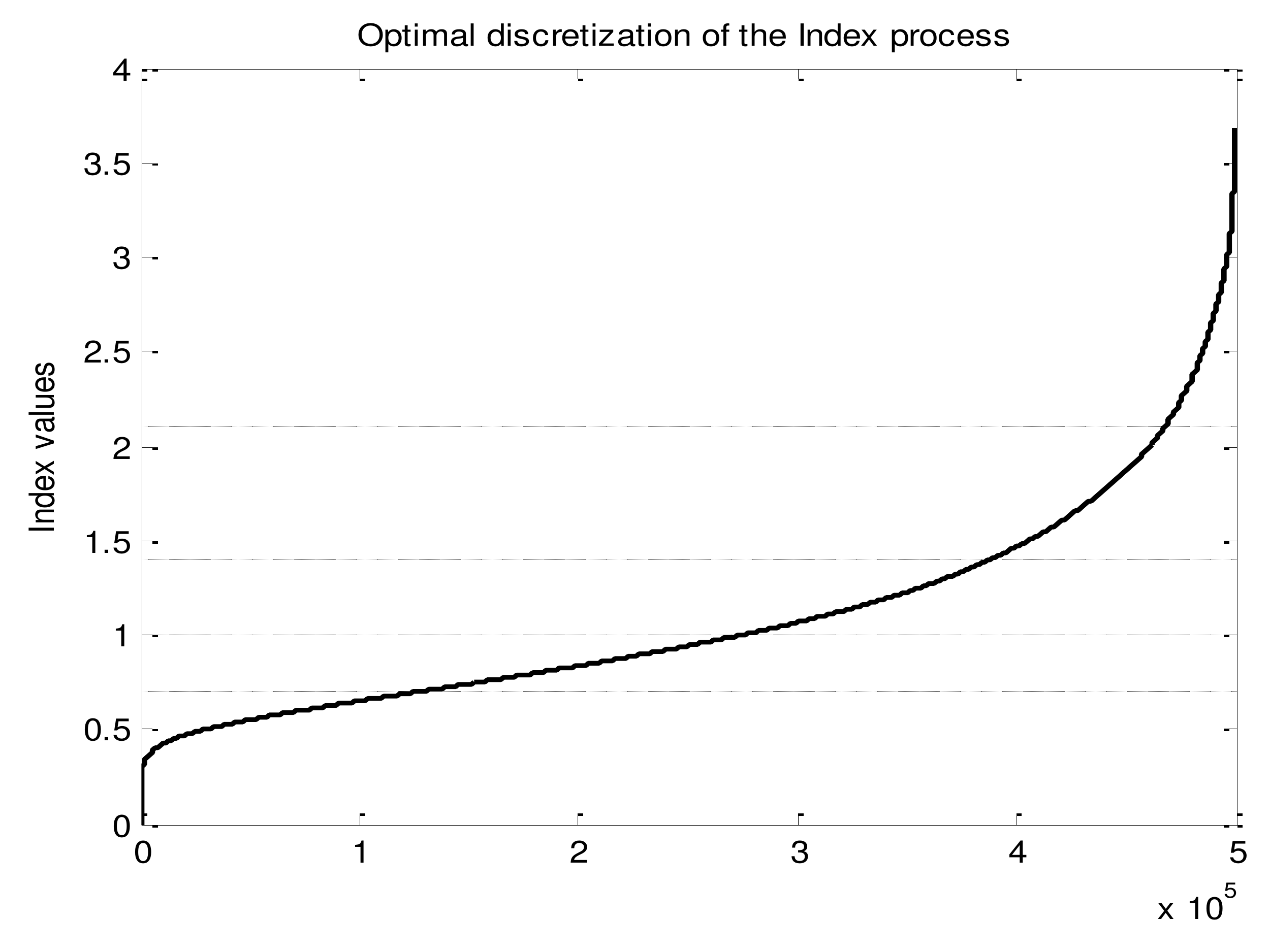}
	\caption{Optimal discretization of the Index process}
	\label{fig:optimal}	
\end{figure}

\begin{table}
	\begin{center}
		\caption{Values of the Index process in case of five change points}
		\begin{tabular}{c|ccccc}
			\label{resfourchp}
				& $\psi_1$ & $\psi_2$ &$ \psi_3$ & $\psi_4$   \\ \hline
			Value & 0.70 & 1.00 & 1.40  & 2.10 \\ 
		\end{tabular} 
	\end{center}
\end{table} 

We identified the optimal number of change points to be four since the BIC index reached an improvement level of less than 0.1\%. As a consequence, the Index process is divided in five states representing five levels of the volatility in the market: very low, medium-low, medium, medium-high and very high.

The border values of the index process obtained in the case of four change points are given in Table \ref{resfourchp} and graphically represented in Figure \ref{fig:optimal} where they are represented jointly with the Index process.  

Once the optimal number of change points is identified, the probability transition matrices of the price return process in each state of volatility can be estimated through the maximum likelihood estimators. The estimated matrices are shown below.

\tiny
\[\hat{P}(1)=
\begin{pmatrix}
0.046 &	0.143 &	0.351 &	0.378 &	0.082 \\
0.014 &	0.141 &	0.445 &	0.374 &	0.026 \\
0.016 &	0.219 &	0.532 &	0.217 &	0.016 \\
0.027 &	0.365 &	0.448 &	0.146 &	0.014 \\
0.084 &	0.360 &	0.358 &	0.147 &	0.051 \\
\end{pmatrix} ; \ 
\hat{P}(2)=
\begin{pmatrix}
0.064 &	0.162 &	0.314 &	0.344 &	0.116 \\
0.033 &	0.202 &	0.375 &	0.340 &	0.050 \\
0.039 &	0.249 &	0.429 &	0.248 &	0.035 \\
0.053 &	0.333 &	0.382 &	0.199 &	0.033 \\
0.107 &	0.343 &	0.321 &	0.169 &	0.060
\end{pmatrix}
\]\\
\[
\hat{P}(3)=
\begin{pmatrix}
0.091 &	0.171 &	0.279 &	0.304 &	0.154 \\
0.067 &	0.215 &	0.321 &	0.308 &	0.089 \\
0.073 &	0.248 &	0.367 &	0.245 &	0.067 \\
0.086 &	0.300 &	0.332 &	0.222 &	0.060 \\
0.135 &	0.312 &	0.283 &	0.182 &	0.088 \\
\end{pmatrix} ; \ 
\hat{P}(4)=
\begin{pmatrix}
0.150 &	0.166 &	0.224 &	0.238 &	0.222 \\
0.127 &	0.199 &	0.269 & 0.256 &	0.149 \\
0.138 &	0.208 &	0.298 &	0.222 &	0.134 \\
0.150 &	0.245 &	0.276 &	0.206 &	0.123 \\
0.206 &	0.236 &	0.245 &	0.172 &	0.141 \\
\end{pmatrix}
\]\\
\[
\hat{P}(5)=
\begin{pmatrix}
0.239 &	0.121 &	0.153 &	0.155 &	0.332 \\
0.232 &	0.151 &	0.182 &	0.169 &	0.266 \\
0.241 &	0.142 &	0.207 &	0.168 &	0.242 \\
0.249 &	0.168 &	0.182 &	0.161 &	0.240 \\
0.320 &	0.150 &	0.155 &	0.130 &	0.245 \\
\end{pmatrix}
\]\\

\normalsize

\normalsize
\begin{table}[h]
	\begin{center}
		\small
		\caption{\% Root square mean deviation: four change points} 	
		\begin{tabular}{c|ccccc}
			\label{rmsd4}	
			& $\hat{P}(1)$ & $\hat{P}(2)$ & $\hat{P}(3)$ & $\hat{P}(4)$& $\hat{P}(5)$ \\ \hline
			$\hat{P}(1)$ & 0.0\% & 20.1\% & 35.8\% & 59.4\% & 100.5\% \\
			$\hat{P}(2)$ & 20.1\% & 0.0\% & 16.8\% & 43.0\% & 86.1\% \\
			$\hat{P}(3)$ & 35.8\% & 16.8\% & 0.0\% & 27.1\% & 70.9\% \\
			$\hat{P}(4)$ & 59.4\% & 43.0\% & 27.1\% & 0.0\% & 44.1\% \\
			$\hat{P}(5)$ & 100.5\% & 86.1\% & 70.9\% & 44.1\% & 0.0\% \\
		\end{tabular} 	
	\end{center}
\end{table} 

\begin{table}[h]
	\begin{center}
		\small
		\caption{\% Mean absolute deviation: four change points} 	
		\begin{tabular}{c|ccccc}
			\label{mad4}	
			& $\hat{P}(1)$ & $\hat{P}(2)$ & $\hat{P}(3)$ & $\hat{P}(4)$& $\hat{P}(5)$ \\ \hline
			$\hat{P}(1)$ & 0.0\% &17.2\% & 32.2\% &53.4\% & 90.1\%  \\
			$\hat{P}(2)$ & 17.2\% & 0.0\% & 15.2\% & 38.6\% &80.6\% \\
			$\hat{P}(3)$ & 32.2\% &15.2\% & 0.0\% & 25.2\%  &  67.7\%\\
			$\hat{P}(4)$ & 53.4\% & 38.6\% & 25.2\%  & 0.0\% & 42.6\% \\
			$\hat{P}(5)$ & 90.1\% & 80.6\% &  67.7\% & 42.6\% & 0.0\% \\
		\end{tabular} 	
	\end{center}
\end{table} 

Noticeable findings can be deduced from these matrices:
\begin{itemize}
	\item $p_{i,3}(h)>p_{i,3}(h+1)$ for all $i\in E$ and $h\in\{1,2,3,4\}$. This inequality expresses the fact that the state 3 (null return) occupancy with next transition has a decreasing probability with respect to the volatility;
	\item For all $i\in E$ and $h\in\{1,2,3,4\}$ $p_{i,5}(h)<p_{i,5}(h+1)$ and $p_{i,1}(h)<p_{i,1}(h+1)$. The inequalities suggest that states characterized by big (negative or positive) returns are more likely to be in presence of high volatility;
	\item For all $h\in\{1,2,3,4,5\}$ and for $i\in \{1,2\}$, $p_{i,1}(h)+p_{i,2}(h)<p_{i,4}(h)+p_{i,5}(h)$. For all $h\in\{1,2,3,4,5\}$ and for $i\in \{4,5\}$, $p_{i,1}(h)+p_{i,2}(h)>p_{i,4}(h)+p_{i,5}(h)$. This means that, independently from the volatility regime ($h\in\{1,2,3,4,5\}$), the occupancy of a low-return state ($i\in \{1,2\}$) increases the probability to reach with next transition a high-return state ($j\in \{4,5\}$) and vice-versa the occupancy of a high-return state ($i\in \{4,5\}$) increases the probability to reach with next transition a low-return state ($j\in \{1,2\}$). Consequently, we can affirm that the return process exhibits a mean reverting property;  
      \item For all $h\in\{1,2,3\}$ $p_{3,1}(h)+p_{3,2}(h)>p_{3,4}(h)+p_{3,5}(h)$, i.e. for low-medium volatility levels from the null-return state ($i=3$) it is more probable to reach with next transition a low-return state ($j\in \{1,2\}$) than a high-return state ($j\in \{4,5\}$). For all $h\in\{4,5\}$ $p_{3,1}(h)+p_{3,2}(h)<p_{3,4}(h)+p_{3,5}(h)$, i.e. for high-volatility levels from the null-return state ($i=3$) it is more probable to reach with next transition an high-return state ($j\in \{4,5\}$) than a low-return state ($j\in \{1,2\}$).   
\end{itemize}

Financial time series present a very important feature: the fact that while the returns are not autocorrelated, 
the square of returns or their absolute values are long rage correlated. It is important that the model describing such dynamics presents the same characteristic.

The autocorrelation of the square of returns for various time lags, which we will denote with $\tau$, is given by the Eq. \ref{autosquare}.

\begin{equation}
\label{autosquare}
\Sigma(\tau)=\frac{Cov(R^2(t+\tau),R^2(t))}{Var(R^2(t))}.
\end{equation}

\begin{figure}[h]
	\centering
	\includegraphics[width=0.8\textwidth]{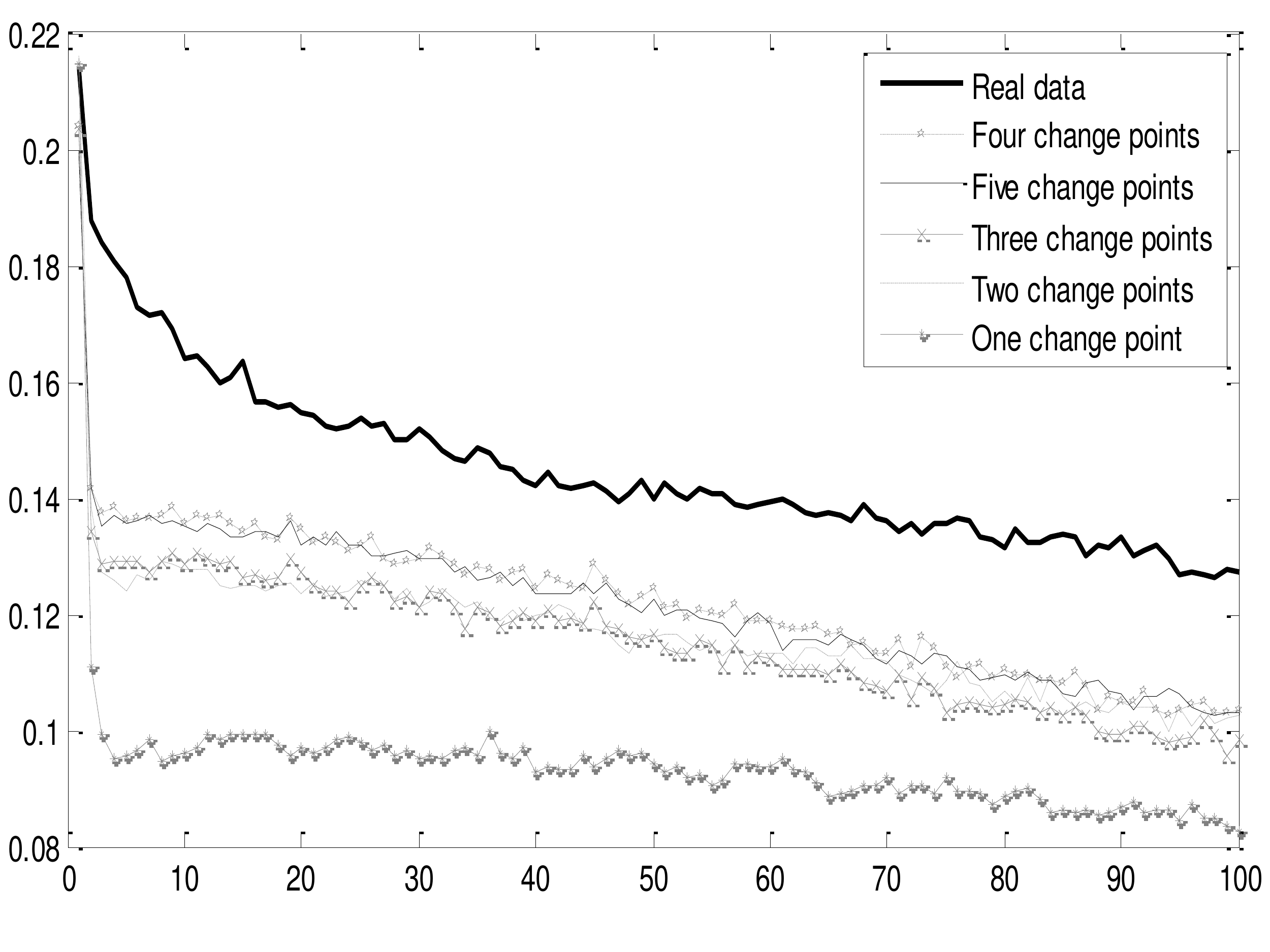}
	\caption{Autocorrelation of the square of returns for various number of change points of the Index process.}
	\label{fig:autocorr}	
\end{figure}

We compared the autocorrelation function of the square of returns of real data with simulated trajectories using the models defined above (see Figure \ref{fig:autocorr})

The graph shows that for the model with one change point, the autocorrelation of the square of returns falls rapidly to zero. In choosing two change points, there is a significant improvement in the long range autocorrelation. In adding more change points, the function continues to improve even though at a very slow rate. It is worth noticing that for the model with four change points the autocorrelation function of the simulated data is the closest to the autocorrelation of the real data. Adding one more change point seems to worsen the result (even though the difference with the model with four change points is minimal).  

\begin{figure}
	\centering
	\includegraphics[width=0.8\textwidth]{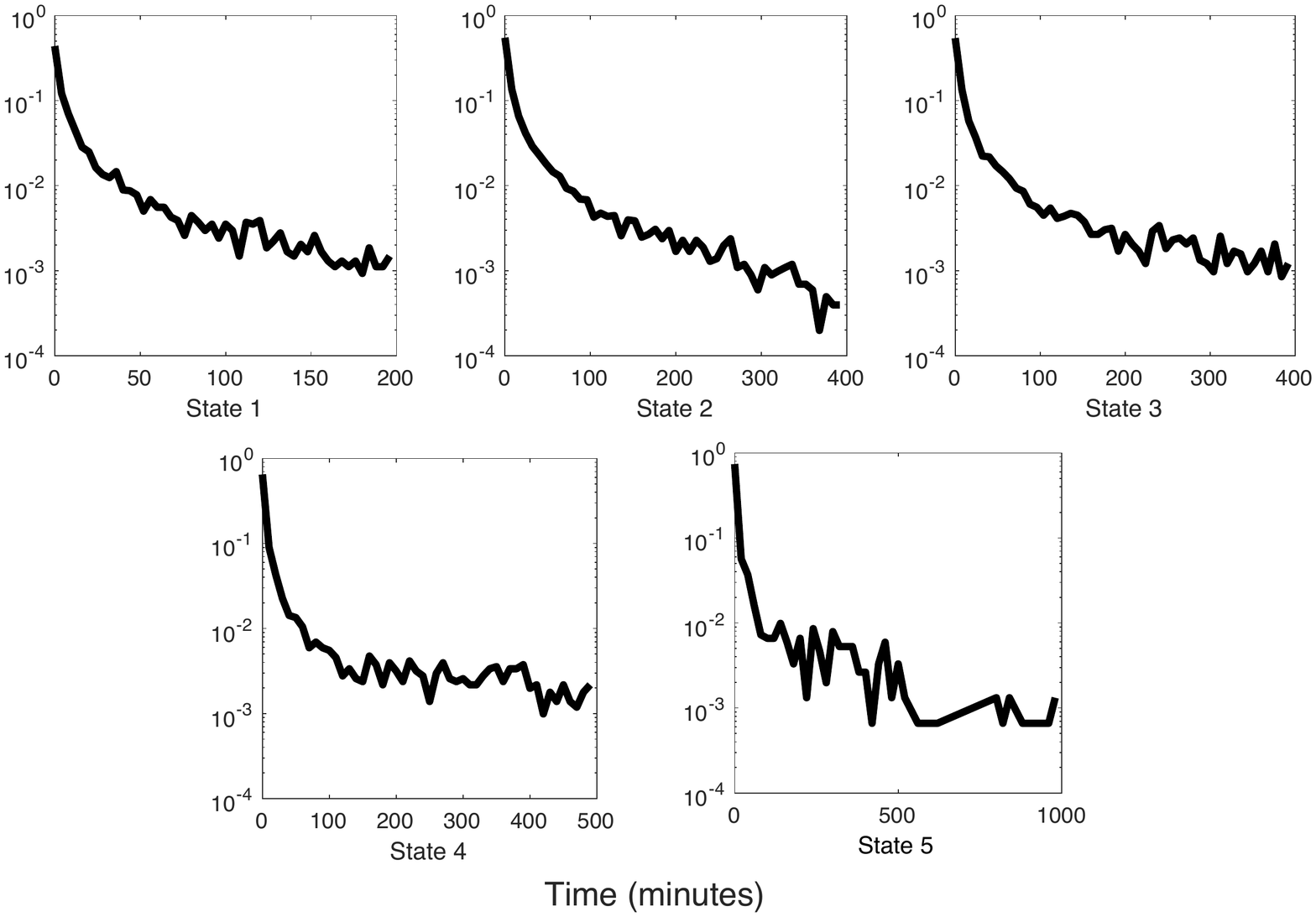}
	\caption{Distributions of time needed to enter in a given volatility state. The distributions have been estimated for each volatility level.}
	\label{fig:fpt}	
\end{figure}

As the last results, in Figure \ref{fig:fpt} we show the distributions of time the volatility process takes to enter in a given state computed using Proposition (\ref{prep1}). Results show different probability distribution of time needed to enter in the different states of volatility (index process) and in general we observe an increase in the time necessary to enter in a volatility state with respect to the value of the volatility. As a matter of example, if we consider the state 1 of volatility we note that the process enters in this state in a time less than $200$ minutes, whereas for state 5 of volatility the entrance in this state occurs in a time less than $1000$ minutes. 

\section{Conclusions}

In this paper we defined an Indexed Markov chain model as a particular case of Indexed-semi-Markov chain models and we confronted with the problem of the determination of the optimal number of states for the index process. We solved the problem by adapting  the change point approach for Markov chains to our more general framework and we calculated the probability distribution function of the first passage time of the index process for different states. The results have been applied to time series of financial returns and stylized facts of financial time series are satisfactorily reproduced by the Indexed Markov chain model. The states of the index process have been interpreted as different regimes of volatility and then our model also furnishes the optimal number of volatility regimes to be used in the valuation of a financial asset.\\ 
\indent Object of future research will be the extension of these techniques to the more general class of indexed-semi-Markov chain models.

\section*{References}
\bibliography{bibl}

\end{document}